\documentclass[11pt]{article}
\usepackage[hmargin=3cm,vmargin=2cm]{geometry}
\usepackage{graphicx, amssymb}
\title{\bf Dirac-Hulth$\acute{e}$n Problem with Position-dependent Mass in $D$-dimensions}
\author{\bf D. Agboola\footnote{e-mail:~tomdavids2k6@yahoo.com}}
\date{\it Department of Mathematics,College of Science and Technology,Covenant University, Ogun State, P.M.B.1023, Nigeria.}
\begin{document}
\maketitle
\vspace{0.5in}
\noindent {\bf Abstract} An approximate solution of the position-dependent mass Dirac equation with the Hulth$\acute{e}$n potential is obtained in $D$-dimensions within frame work of an exponential approximation of the centrifugal term. The relativistic energy spectrum is worked out using direct transformation method; the two-component spinor wavefunctions are obtained in terms of the Jacobi polynomials. Dependence of the energy levels on some parameters is discussed. The results obtained are in good agreement with previous works. 

\maketitle
\vspace{0.5in}
\vspace{.5in}


\maketitle

\section{ Introduction}
The relativistic Dirac equation plays an important role in the understanding of many quantum mechanical problems. The equation which describe the motion of a spin-$\frac{1}{2}$ particle has been solved with various physical potentials. For instance, the solutions of the Dirac equation with the Coulomb potential have been obtained in 1$D$ [1, 2], 2$D$ [3] and 3$D$ [4-6]. Also, the qualitative properties of the Dirac particle in a central potential and the discrete eigenvalues of the radial Dirac operator have been discussed [7, 8]. However, with the intreest in the higher dimensional field theory, the multidimensional relativistic and non-relaivistic equations have been studied by many authors. To mention a few, the $D$-dimensional Schr\"{o}dinger has been studied with the Coulomb potential [9], pseudoharmonic potential [10], Hulth$\acute{e}$n potential [11] and P\"oschl-Teller potential [12]. Also, various potentials have been studied with both the $D$-dimensional klein-Gordon [13-16] and Dirac [17-20] equations.

Moreover, the Hulth$\acute{e}$n potential is one of the important short-range potentials
which behaves like a Coulomb potential for small values of $r$ and decreases
exponentially for large values of $r$. The Hulth$\acute{e}$n potential has received extensive study in both relativistic and non-relativistic quantum mechanics [11, 13-16, 21, 22]. Unfortunately, quantum mechanical equations with the Hulth$\acute{e}$n potential can be solved analytically only for the $s$-states [23-25]. However, recent study [26] has used an exponential approximation for the centrifugal term in order to obtain the solution of the $\ell\neq 0$ states of the Hulth$\acute{e}$n potential. This approximation has been empolyed to obtain the solutions of both the relativistic [14] and non-relativistic [11] Hulth$\acute{e}$n potential in $D$-dimensions. On the other hand, the used of position-dependent mass has produced interesting results in many bound-state problems. This approach has been used by many authors which include [27-30].  The present study is of two-fold. The first purpose is to study the bound state of the Dirac-Hulth$\acute{e}$n problem in $D$-dimensions. Secondly, we to investigate if the dimensional degeneracy observed in the non-relativistic Hulth$\acute{e}$nspectrum [11] also occur in the case of the relativistic equations. 

The paper is organized as follows. In section 2, we present the Dirac-Hulth$\acute{e}$n problem in $D$-dimensions. The solutions to the two-component wavefunctions and the relativistic energy spectrum are obtain in section 3. In section 4, the dimensional degeneracy for both Klein-Gordon and the Dirac equations are discussed and some concluding remarks are given in section 5.

\section{The Dirac equation in $D$-dimensions}
The $D$-dimensional Dirac equation  with a central potential $V(r)$ and position-dependent mass $\mu(r)$ can be written in natural units $\hbar=c=1$ as [18, 31]
$$H\Psi{(r)}=E_{n_r\kappa}\Psi(r)\hspace{.2in}\mbox{where}\hspace{.1in}H=\sum_{j=1}^D\hat{\alpha}_jp_j+\hat{\beta}\mu(r)+V(r) \eqno{(1)}$$ where $E_{n\kappa}$ is the relativistic energy, $\{\hat{\alpha}_j\}$ and $\hat{\beta}$ are Dirac matrices, which satisfy anti-commutation relations
$$\begin{array}{lrl}
\hat{\alpha}_j\hat{\alpha}_k+\hat{\alpha}_k\hat{\alpha}_j&=&2\delta_{jk}\bf{1}\\
\hat{\alpha}_j\hat{\beta}+\hat{\beta}\hat{\alpha}_j&=&0\\
{\hat{\alpha}_j}^2=\hat{\beta}^2&=&\bf{1}

\end{array} \eqno{(2)}$$
and 
$$p_j=-i\partial_j=-i\frac{\partial}{\partial x_j} \hspace{.2in} 1\leqslant j\leqslant D. \eqno{(3)}$$ The orbital angular momentum operators $L_{jk}$, the spinor opertaors $S_{jk}$ and the total angular momentum operators $J_{jk}$ can be defined as follows:
$$L_{jk}=-L_{jk}=ix_j\frac{\partial}{\partial x_k}-ix_k\frac{\partial}{\partial x_j},\hspace{.2in} S_{jk}=-S_{kj}=i\hat{\alpha}_j\hat{\alpha}_k/2,\hspace{.2in} J_{jk}=L_{jk}+S_{jk}.$$
$$L^2=\sum_{j<k}^DL^2_{jk},\hspace{.2in}S^2=\sum_{j<k}^DS^2_{jk},\hspace{.2in}J^2=\sum_{j<k}^DJ^2_{jk}, \hspace{.2in} 1\leqslant j< k\leqslant D. \eqno{(4)}$$
For a spherically symmetric potential, total angular momentum operator $J_{jk}$ and the spin-orbit operator $\hat{K}=-\hat{\beta}(J^2-L^2-S^2+(D-1)/2)$ commutate with the Dirac Hamiltonian. For a given total angular momentum $j$, the eigenvalues of $\hat{K}$ are $\kappa=\pm(j+(D-2)/2)$; $\kappa=-(j+(D-2)/2)$ for aligned spin $j=\ell+\frac{1}{2}$ and $\kappa=(j+(D-2)/2)$ for unaligned spin $j=\ell-\frac{1}{2}$. Also, since $V(r)$ is spherically symmetric, the symmetry group of the system is SO$(D)$ group.

Thus, we can introduce the hyperspherical coordinates [32-35]
$$\begin{array}{lrl}
x_1&=&r\cos\theta_1\\
x_\alpha&=&r\sin\theta_1\dots\sin\theta_{\alpha-1}\cos\phi,\hspace{.2in}2\leqslant \alpha\leqslant D-1\\
x_D&=&r\sin\theta_1\dots\sin\theta_{D-2}\sin\phi,\\
\end{array} \eqno{(5)}$$ where the volume element of the configuration space is given as
$$\prod_{j=1}^Ddx_j=r^{D-1}drd\Omega \hspace{.2in} d\Omega=\prod_{j=1}^{D-1}(\sin\theta_j)^{j-1}d\theta_j  \eqno{(6)}$$
with $0\leqslant r< \infty$, \hspace{.1in}$0\leqslant\theta_k\leqslant\pi$, $k=1,2,\dots D-2$,\hspace{.1in} $0\leqslant\phi\leqslant 2\pi$, such that the spinor wavefunctions can be classified according to the hyperradial quantum number $n_r$ and the spin-orbit quantum number $\kappa$ and can be written using the Pauli-Dirac representation 
$$\Psi_{n_r\kappa}(r,\Omega_D)=r^{-\frac{D-1}{2}}\left(\begin{array}{lll}
F_{n_r\kappa}(r)Y_{jm}^\ell\left(\Omega_{D}\right)\\\\
iG_{n_r\kappa}(r)Y^{\tilde{\ell}}_{jm}\left(\Omega_{D}\right)
\end{array}\right) \eqno{(7)}$$ 
where $F_{n_r\kappa}(r)$ and $G_{n_r\kappa}(r)$ are the radial wave function of the upper- and the lower-spinor components respectively, $Y_{jm}^\ell\left(\Omega_{D}\right)$ and $Y^{\tilde{\ell}}_{jm}\left(\Omega_{D}\right)$ are the hyperspherical harmonic functions coupled with the total angular momentum $j$. The orbital and the pseudo-orbital angular momentum quantum numbers for spin symmetry  $\ell$ and  and pseudospin symmetry $\tilde{\ell}$ refer to the upper- and lower-component respectively.  

Substituting Eq.\,(7) into Eq.\,(1), and seperating the variables we obtain the following coupled radial Dirac equation for the spinor components:
$$\left(\frac{d}{dr}+\frac{\kappa}{r}\right)F_{n_r\kappa}(r)=[\mu(r)+E_{n_r\kappa}-V(r)]G_{n_r\kappa}(r)\eqno{(8)}$$
$$\left(\frac{d}{dr}-\frac{\kappa}{r}\right)G_{n_r\kappa}(r)=[\mu(r)-E_{n_r\kappa}+V(r)]F_{n_r\kappa}(r)\eqno{(9)}$$
where $\kappa=\pm(2\ell+D-1)/2$. Further details of the derivation can be obtain from refs [36-38]. Using Eq.\,(8) as the upper component and substituting into Eq.\,(9), we obtain the follwoing second order differential equations
$$\left[\frac{d^2}{dr^2}-\frac{\kappa(\kappa+1)}{r^2}-[\mu(r)+E_{n_r\kappa}-V(r)][\mu(r)-E_{n_r\kappa}+V(r)]-\frac{\left(\frac{d\mu(r)}{dr}-\frac{dV(r)}{dr}\right)\left(\frac{d}{dr}+\frac{\kappa}{r}\right)}{[\mu(r)+E_{n_r\kappa}-V(r)]}\right]F_{n_r\kappa}(r)=0 \eqno{(10)}$$
$$\left[\frac{d^2}{dr^2}-\frac{\kappa(\kappa-1)}{r^2}-[\mu(r)+E_{n_r\kappa}-V(r)][\mu(r)-E_{n_r\kappa}+V(r)]-\frac{\left(\frac{d\mu(r)}{dr}+\frac{dV(r)}{dr}\right)\left(\frac{d}{dr}-\frac{\kappa}{r}\right)}{[\mu(r)-E_{n_r\kappa}+V(r)]}\right]G_{n_r\kappa}(r)=0 \eqno{(11)}$$
We note that the energy eigenvalues in these equation depend on the angular momentum quantum number $\ell$ and dimension $D$. However, to solve these equations, we shall use an approximation for the centrifugal barrier as discussed in the following section.

\section{Bound states of the Dirac-Hulth$\acute{e}$n problem in $D$-dimension}
We start this section by defining the Hulth$\acute{e}$n potential as follows [11, 13, 14, 21, 22]
$$V(r)=-Z\alpha\frac{e^{-\alpha r}}{1-e^{-\alpha r}}  \eqno{(12)}$$
where $\alpha$ is the screening parameter and $Z$ is a constant which is identified with the atomic number when the potential is used for atomic phenomenon. To solve Eq.\,(10), we first eliminate the last term by equating $\frac{d\mu(r)}{dr}-\frac{dV(r)}{dr}=0$, which gives the mass function 
$$\mu(r)=\mu_0+\frac{Z\alpha}{1-e^{-\alpha r}} \eqno{(13)}$$ where $\mu_0$ is the intergation constant, and approximate the centrifugal term as follows [11, 26]
$$\frac{1}{r^2}\approx \frac{\alpha^2 e^{-\alpha r}}{(1-e^{-\alpha r})^2}.  \eqno{(14)}$$
Substituting Eqs.\,(12), (13) and (14) into Eq.\,(10) we have 
$$\left[\frac{d^2}{dr^2}-\frac{\kappa(\kappa+1)\alpha^2 e^{-\alpha r}}{(1-e^{-\alpha r})^2}-\left(\mu_0+\frac{Z\alpha}{1-e^{-\alpha r}}+E_{n_r\kappa}+Z\alpha\frac{e^{-\alpha r}}{1-e^{-\alpha r}} \right)\right.$$
$$\left.\times\left(\mu_0+\frac{Z\alpha}{1-e^{-\alpha r}}-E_{n_r\kappa}-Z\alpha\frac{e^{-\alpha r}}{1-e^{-\alpha r}}\right)\right]F_{n_r\kappa}(r)=0\eqno{(15)}$$
Taking the transformation $s=e^{-\alpha r}$, Eq.\,(15) becomes 
$$\left[\frac{d^2}{ds^2}+\frac{1}{s}\frac{d}{ds}-\frac{\kappa(\kappa+1)}{s(1-s)^2}-\frac{\epsilon^2}{s^2}-\frac{\beta_1+\beta_2}{s(1-s)}\right]F_{n_r\kappa}(s)=0 \eqno{(16)}$$ where 
$$\epsilon=\frac{\sqrt{\mu_0^2+\beta_1\alpha^2-E_{n_r\kappa}^2}}{\alpha},\hspace{.1in}\beta_1=\frac{2Z\mu_0+Z^2\alpha}{\alpha}\hspace{.1in}\mbox{and} \hspace{.1in}\beta_2=\frac{2ZE_{n_r\kappa}+Z^2\alpha}{\alpha}\eqno{(17)}$$
If we seek the solution of the form
$$F_{n_r\kappa}(s)=s^\epsilon(1-s)^{\delta}U_{n_r\kappa}(s), \eqno{(18)}$$ then Eq.\,(16) becomes
$$ U_{n_r\kappa}''(s)+\left(\frac{1+2\epsilon}{s}-\frac{2\delta}{1-s}\right)U'_{n_r\kappa}(s)+$$
$$\left[\frac{\left(-(\beta_1+\beta_2)-(2\epsilon+1)\delta-\delta^2+\delta\right)(1-s)+\delta^2-\delta-\kappa(\kappa+1)}{s(1-s)^2}\right]U_{n_r\kappa}(s)=0 \eqno{(19)}$$
One can choose $\delta^2-\delta-\kappa(\kappa+1)=0, \hspace{.1in}\mbox{i.e}\hspace{.1in} \delta=\kappa+1$~(the positive root), such that Eq.\,(19) becomes
$$s(1-s)U''_{n_r\kappa}(s)+\left[(1+2\epsilon)-(2\epsilon+2\delta+1)s)\right]U_{n_r\kappa}'(s)$$
$$-\left[\delta+\epsilon-\sqrt{\epsilon^2-(\beta_1+\beta_2)}\right]\left[\delta+\epsilon+\sqrt{\epsilon^2-(\beta_1+\beta_2)}\right]U_{n_r\kappa}(s)=0. \eqno{(20)}$$
Eq.\,(20) is the well-known hypergeometric equation whose solution is given in form of the hypergeometric function [39] 
$$U_{n_r\kappa}(s)=\ _2F_1\left[\delta+\epsilon-\sqrt{\epsilon^2-(\beta_1+\beta_2)},~\delta+\epsilon+\sqrt{\epsilon^2-(\beta_1+\beta_2)};~ 1+2\epsilon;~s\right].\eqno{(21)}$$
However, for large value of $s$, the solution in (20) diverges, thus preventing normalization. To avoid this, we set 
$$-n_r=\delta+\epsilon-\sqrt{\epsilon^2-(\beta_1+\beta_2)},\hspace{.15in}n_r=0,1,2,\dots.\eqno{(22)}$$
from which, with the help of Eq.\,(17) we have the energy eigenvalues  
$$E_{n_r\kappa}=-\frac{Z\alpha(\eta+\beta_1)}{2\eta}\pm\frac{(n_r+|\kappa|+1)}{2\eta }\sqrt{4(\mu_0+\alpha^2\beta_1)\eta-\alpha^2(\eta+\beta_1)^2}\eqno{(23)}$$
where $\eta=(n_r+|\kappa|+1)^2+Z^2$. To check the validity of the energy spectrum, we take the limit as $\alpha\rightarrow 0$ of Eq.\,(23) and this yields
$$E_{Col}=\sqrt{\mu_{0}}\left[1+\frac{Z^2}{(n_r+|\kappa|+1)^2}\right]^{-\frac{1}{2}}  \eqno{(24)}$$ which is the energy spectrum of the Coulomb-like potential in $D$-dimensions [17, 36]. Evidently, this follows from the fact that $\lim_{\alpha\rightarrow 0}V(r)=-\frac{Z}{r}$.

Thus the unnormalized wavefunction for the upper-component can be written as
$$F_{n_r\kappa}(s)=C_{n_r\kappa}s^{\epsilon}(1-s)^{\kappa+1}\,_2F_1\left[-n_r,~\kappa+\epsilon+1+\sqrt{\epsilon^2-(\beta_1+\beta_2)};~ 1+2\epsilon;~s\right] \eqno{(25)}$$ Using the relation
using the following definition of the Jacobi polynomial [39]
$$P^{(a,b)}_n(s)=\frac{\Gamma(n+a+1)}{n!\Gamma(1+a)} \ _2F_1\left(-n,a+b+n+1;1+a;\frac{1-s}{2}\right),  \eqno{(26)}$$ we arrive at
$$F_{n_r\kappa}(s)=C_{n_r\kappa}s^{\epsilon}(1-s)^{\kappa+1} P_{n_r}^{(2\epsilon,~2\kappa+1)}(1-2s) \eqno{(27)}$$
 where $C_{n_r\kappa}$ is the normalization constant.

Moreover, substituting Eq.\,(27) into Eq.\,(8), one can easily obtain the lower-component as 
$$G_{n\kappa}(s)=A_1(s)P_{n_r}^{(2\epsilon,~2\kappa+1)}(1-2s)+ A_2(s)P_{n_r-1}^{(2\epsilon+1,~2\kappa+2)}(1-2s) \eqno{(28)}$$ where
$$A_1(s)=\frac{C_{n_r\kappa}s^\epsilon(1-s)^\kappa\left[{\epsilon}/{s}-{\alpha\kappa(1-s)}/{\log s}\right]}{\mu(r)+E_{n_r\kappa}-V(r)}\hspace{.1in} \mbox{and} \hspace{.1in} A_2(s)=\frac{B_{n_r\kappa}s^\epsilon(1-s)^\kappa}{\mu(r)+E_{n_r\kappa}-V(r)} \eqno{(29)}$$ with $B_{n_r\kappa}$ a constant.

\section{Dependence of the relativistic Hulth$\acute{e}$n spectra on some\\ parameters}
\subsection{Case 1: Dependence of $E_{n}$ on $\alpha$}
Recently, it was reported in [11] that the non-relativistic energy spectrum of the Hulth$\acute{e}$n potential intersect for some adjacent dimensions $D (\geq 3)$, for some  values of the screening parameter $\alpha_i$. Thus signifying a degeneracy (with respect to the dimension) in the energy spectrum of the non-relativistic Hulth$\acute{e}$n potential. In this section, we investigate if such degeneracy also occur in the relativistic energy spectums of the Hulth$\acute{e}$n potential. We start with the Klein-Gordon-Hulth$\acute{e}$n energy spectrum for equal vector and scalar potentials, which is given as [11, 13, 14]   
$$E_{K_{n_r}}=-\frac{Z\gamma}{2\eta}\pm\frac{(n_r+v_1)^2}{2\eta}\sqrt{4\mu^2-\alpha(2Z\mu-\gamma)}\eqno{(30)}$$
 where $n_r$ is the radial quantum number, $\mu$ is the mass and $$v_1=(D+2\ell-1)/2,\hspace{0.1in}\gamma={2Z\mu}-\alpha(n_r+v_1)^2 \hspace{0.1in}\mbox{and}\hspace{0.1in} \eta=(n_r+v_1)^2+Z^2. \eqno{(31)}$$
Moreover, if we set $Z=\mu=1$, and define a principal quantum number $n=n_r+\ell +1$, the energy eqution (30) takes the simple form (with the choice of a negative root)
$$E_{K_{n}}=\frac{2\alpha\rho_1-16}{4+\rho_1}-\frac{\rho_1}{4+\rho_1}\sqrt{16-\alpha^2\rho_1}.\eqno{(32)}$$ where $\rho_1=(2n+D-3)^2$. Similar to result of [11], the curve of $E_{K_n}$ against $\alpha$ (Figure 1) reveals the intersection of the energy levels for some adjacent dimensions. Thus indicating degeneracy (with respect to the dimensions) in the energy levels.

We now consider the case of the Dirac-Hulth$\acute{e}$n spectrum. If we introduce a principal quantum number $n=n_r+|\kappa|+1$, the energy equation (23) can be written ($Z=\mu_0=1$) in the form 
$$E_{D_n}=-\frac{\alpha(\rho_2+1)+(\alpha+2)}{2(\rho_2+1)}-\frac{\rho_2}{2(\rho_2+1)}\sqrt{4(\rho_2+1)(\alpha+1)^2-\left[\alpha(\rho_2+2)+2\right]^2} \eqno{(33)}$$
where $\rho_2=(n+(D-1)/2)^2$. In figure 2, the cuvres of $E_{D_n}$ agaisnt $\alpha$ reveals that the Dirac-Hulth$\acute{e}$n spectrums are more evenly distributed with respect to the spatial dimensions; moreover, energy levels at higher exited states, tend to be bounded within [-1,~0]. 

Moreover, for large $\alpha$, the energy eigenvalues (30) and (33) becomes imaginary. In particular if $\alpha>\frac{4}{2n+D-3}$, the Klein-Gordon energy (30) becomes imaginary for the state $E_{K_n}(n,\alpha,D)$. Similarly, if $\alpha> 2\left(1+\sqrt{1+\rho_2}\right)/{\rho_2}$, where  $\rho_2=(n+(D-1)/2)^2$, then the Dirac energy (33) becomes imaginary for the state $E_{D_n}(n,\alpha,D)$.
\subsection{Case 2 : Dependence of $E_n$ on $D$}
To investigate the dependence of the relativistic Hulth$\acute{e}$n energies on the dimension, we follow a recent study [40, 41] where the property of {\it continuous} dimension is used in studing the bound states of a quantum system with special potential. Thus, we assume the dimension $D$ is continuous and plot some curves of $E_n$ against $\alpha$ as shown in figures 3 and 4 for the cases of Klein-Gordon and Dirac spectrum respectively.

It is interesting to note that the set of points in figure 3 indicates that the Klein-Gordon-Hulth$\acute{e}$n energy preserves its discrete nature regardless of the the nature of $D$ and the value of $\alpha$. This is evident from the fact that in Eq.\,(32), if $\alpha$ is sufficiently small and $D$ is sufficiently large, the energy $E_{K_n}$ tends to -4, thus becoming independent of $D$ for any given value of $n$. On the other hand, figure 4 reveals that the Dirac-Hulth$\acute{e}$n spectrum decreases and converges to a point with increase in the dimension and also for large $\alpha$, there is a posssiblity of degenerate energy with respect to $n$.

\section{Concluding remarks}
We have studied the $D$-dimensional Dirac equation with the Hulth$\acute{e}$n potential within the frame work of an exponential approximation of the centrifugal term and position-dependent mass. The approximate eigenvalues obtained were found to be in good agreement with previous works [17]; the eigenfunctions are also obtained in form of the Jacobi polynomials. The dependence of the energy spectrum on the screening parameter $\alpha$ and dimensions $D$ were studied for both the $D$-dimensional Klein-Gordon- and the Dirac-equations with the Hulth$\acute{e}$n potential. We note that similar to the non-relativistic case, the Klein-Gordon-Hulth$\acute{e}$n energy levels for some adjacent dimensions intersect at some values of $\alpha$, thus resulting into a degeneracy of energy eigenvalues with respect to the dimensions. However, the Dirac-Hulth$\acute{e}$n spectrums are more evenly distributed with respect to $\alpha$ and $D$. Finally, we observed that because of the non-linear terms in the energy equations (32) and (33), not all values of $\alpha$ yield a real eigenvalues for a given quantum state. 

\section*{Acknowledgements}
DA is grateful to Prof. S. H. Dong for communicating some of his works during the preparation of the manuscript.

\pagebreak

\begin{center}
   \bf FIGURES
\end{center}

\begin{figure}[h]
\includegraphics[scale=.6]{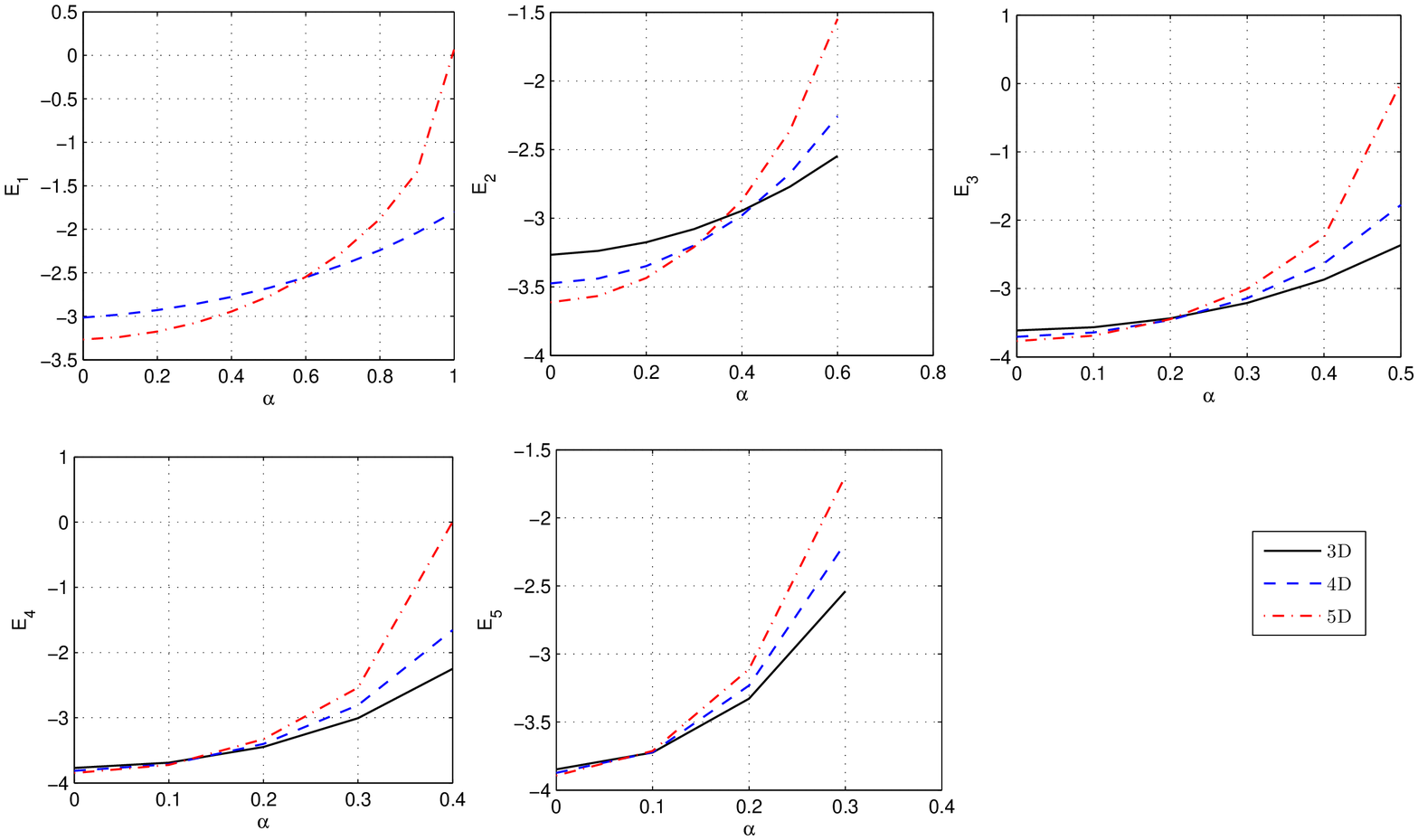}
\caption{The curves of the Klien-Gordon-Hulth$\acute{e}$n energy spectrum against $\alpha$ for some dimensions and some exited states.}
\label{fig:}
\end{figure}

\begin{figure}[h]
\includegraphics[scale=.6]{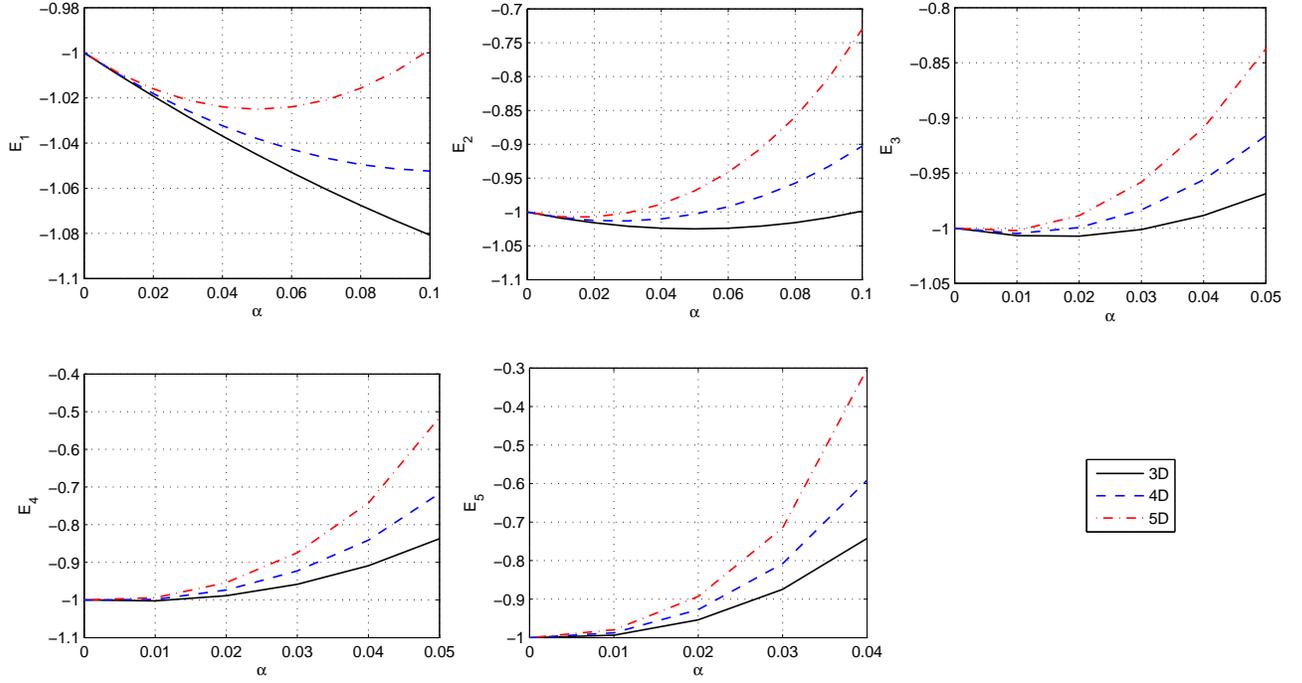}
\caption{The curves of the Dirac- Hulth$\acute{e}$n energy spectrum against $\alpha$ for some dimensions and some exited states.}
\label{fig:}
\end{figure}

\begin{figure}[h]
\includegraphics[scale=.6]{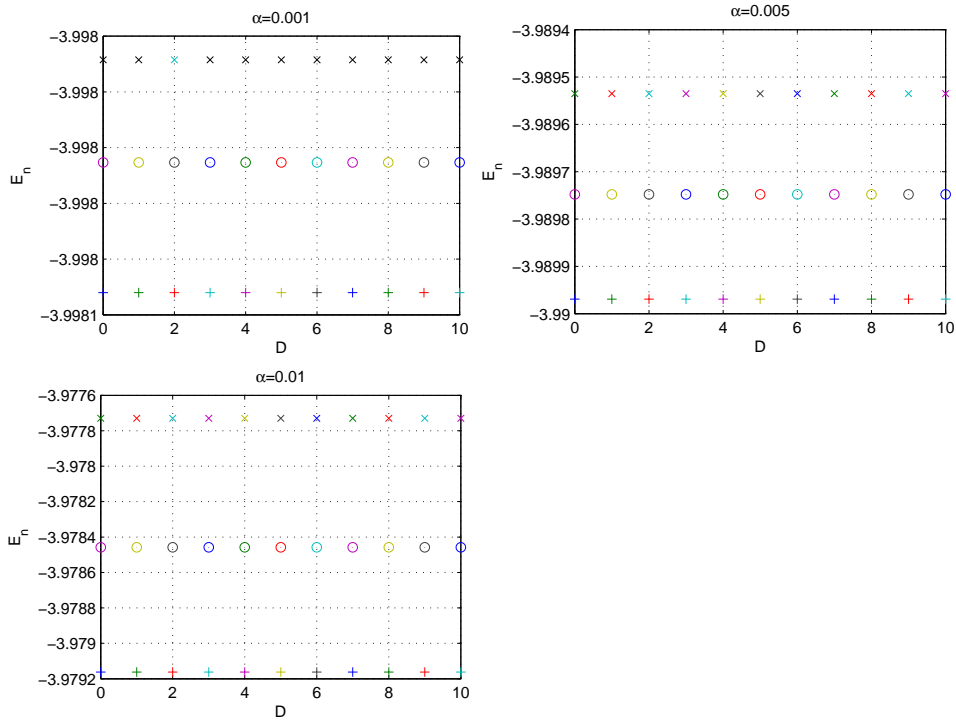}
\caption{The curves of the Klein-Gordon-Hulth$\acute{e}$n energy spectrum against $D$ for some exited states. The points '+', 'o' and 'x' denote $n=3, 4$ and $5$ repectively}
\label{fig:}
\end{figure}

\begin{figure}[h]
\includegraphics[scale=.6]{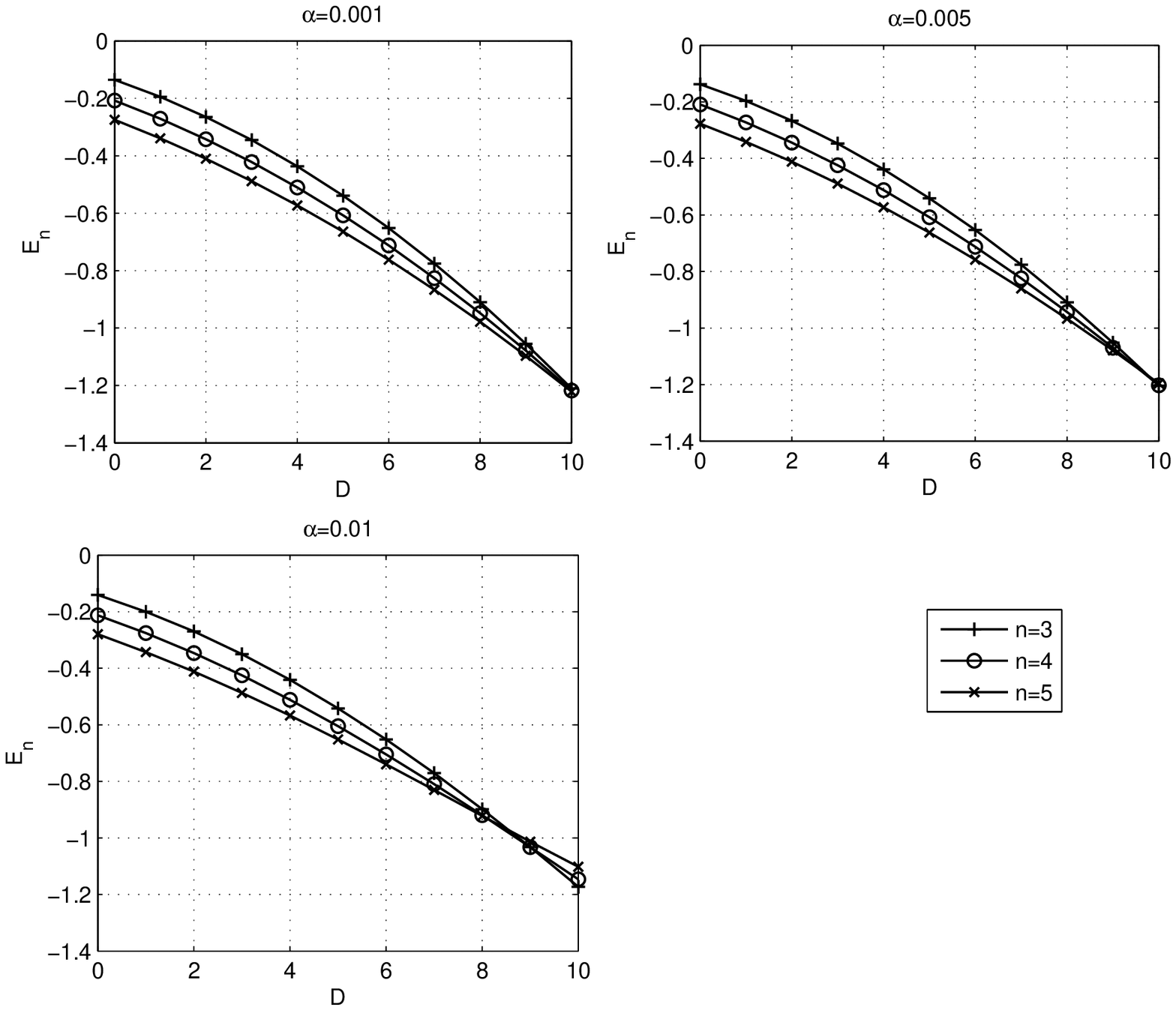}
\caption{The curves of the Dirac-Hulth$\acute{e}$n energy spectrum against $D$ for some exited states.}
\label{fig:}
\end{figure}

\end{document}